\documentclass[pdflatex,sn-mathphys-num]{sn-jnl} 

\usepackage{graphicx}
\usepackage{dcolumn}
\usepackage{bm}
\usepackage{printlen}
\usepackage{lipsum}
\usepackage{siunitx}
\DeclareSIUnit{\pixel}{px}
\DeclareSIUnit{\fps}{fps}
\usepackage{soul}
\usepackage{amsmath,amssymb}

\newcommand{\Ste}{\textit{Ste}}
\newcommand{\Nu}{\textit{Nu}}

\newcommand{\Pra}{\textit{Pr}}
\newcommand{\Rey}{\textit{Re}}
\newcommand{\bnhat}{\mathbf{\hat{n}}}

\def\be{\begin{equation}}
\def\ee{\end{equation}}
\def\ba{\begin{eqnarray}}
\def\ea{\end{eqnarray}}

\usepackage{xcolor}

\begin{document}
\title{\bf Transition from classical to ultimate melting}

\author*[1]{\fnm{Edoardo} \sur{Bellincioni}}\email{e.bellincioni@utwente.nl}
\equalcont{These authors contributed equally to this work.}

\author*[1]{\fnm{Kevin} \sur{Zhong}}\email{k.zhong@utwente.nl}
\equalcont{These authors contributed equally to this work.}

\author[2]{\fnm{Christopher J. } \sur{Howland}}

\author[5]{\fnm{Yiyu}\sur{Zhou}}

\author[1]{\fnm{Sander G.} \sur{Huisman}}

\author[1,3]{\fnm{Roberto} \sur{Verzicco}}

\author*[1,4]{\fnm{Detlef} \sur{Lohse}}\email{d.lohse@utwente.nl}

\affil*[1]{\orgdiv{Physics of Fluids Department and Max Planck Center for Complex Fluid Dynamics and J. M. Burgers Centre for Fluid Dynamics}, \orgname{University of Twente}, \orgaddress{\street{P. O. Box 217}, \city{Enschede}, \postcode{7500 AE}, \country{The Netherlands}}}

\affil[2]{\orgdiv{School of Mathematics and Statistics}, \orgname{University College Dublin}, \orgaddress{\street{Belfield}, \city{Dublin 4}, \country{Ireland}}}

\affil[3]{\orgdiv{Gran Sasso Science Institute}, \orgaddress{\street{Viale Francesco Crispi, 7}, \city{L'Aquila}, \postcode{67100}, \country{Italy}}}

\affil[4]{\orgdiv{Max Planck Institute for Dynamics and Self-Organisation}, \orgaddress{\street{Am Fa{\ss}berg 17}, \city{G\"{o}ttingen}, \postcode{37077}, \country{Germany}}}

\affil[5]{\orgdiv{Department of Modern Mechanics},\orgname{University of Science and Technology of China}, \orgaddress{\city{Hefei}, \postcode{230027}, \country{China}}}

\date{\today}

\newpage

\abstract{
Melting is omnipresent in nature and technology. Its applications range from metallurgy \cite{met11081297,chakraborty2009,aboutalebi1995coupled}, 
biology \cite{farrant1965,uhrig2022}, food science \cite{mathijssen2023culinary,norton2006computational}, and latent thermal energy storage \cite{vogel2019,li2019review,jouhara2020}
to oceanography, geophysics, and climate science \cite{hewitt2020,straneo2015,cenedese2023,pritchard2012,shepherd2012,stroeve2007,sutherland2019,buzzard2022,malyarenko2020,young2022,larter2022,dauxois2021,noel2023,hanna2024,lucas2025}. Furthermore, melting occurs on all scales from sub-millimeter to global scales.
The key objective is to understand the rate at which an object melts as a function of its size and of the ambient conditions. 
To achieve this it is important to be able to extrapolate from small scale experiments and observations to large or even global scales. 
This is done by scaling laws, which are only meaningful if there is no transition from one scaling relation to another one. 
Here we show that such a transition does exist for melting in turbulent flow.  Namely, from slow melting at the small scales to fast melting at the large scales, for both fixed and freely-advected melting objects.
We do so by controlled melting experiments and corresponding direct numerical simulations, covering four orders of magnitude in scale. The transition corresponds to the transition from a laminar-type boundary layer around the melting object to a turbulent-type boundary layer, i.e., from so-called classical turbulence to ultimate turbulence, with its enhanced transport properties \cite{kra62,avi23,lohse2023,lohse2024}. 
Our results thus provide a quantitative understanding of the flow physics of the melting process and thereby enable a better extrapolation and prediction of melt rates on large scales such as relevant in geophysics, oceanography, and climate science \cite{hewitt2020,straneo2015,cenedese2023,hester2021,malyarenko2020,wells_geophysical-scale_2008,pritchard2012,shepherd2012,noel2023,hanna2024,lucas2025}. 
}

\maketitle

\section*{Introduction}
The melting of icebergs or glacial ice into the ocean is not understood on a quantitative level. Such an understanding, however, is essential for reliable ocean and climate models and to predict the rise of the sea level. A number of issues are identified in literature: 
The melt rate predictions from present models, which are based on effective thermal diffusivities, 
  are often off by an order of magnitude or even more as compared to recent field measurement \cite{stroeve2007,sutherland2019,buzzard2022}; 
  the inter-model variability is significant \cite{malyarenko2020}; and basal glacier melting into the ocean plays a much more prominent role than previously thought \cite{young2022,larter2022}.The lack of understanding is on a fundamental level and one of the grand challenges in environmental fluid dynamics \cite{dauxois2021}; ice melting is a highly complex multi-scale, multi-physics problem, with multi-way coupling between heat and salt transport, nonlinear buoyancy effects, fluid flow, and an evolving ice-water interface \cite{cenedese2023,hewitt2020,weady2022,rubinstein1971}. What contributes to the complexity of that flow is, in many circumstances, turbulence around
melting ice \citep{malyarenko2020,cenedese2023}, which in itself is already a
multi-scale phenomenon.
 
A fundamental understanding of melting 
processes in turbulent flow is hugely relevant also in various other processes in nature and technology, ranging from metallurgy \cite{met11081297,aboutalebi1995coupled}, 
food science \cite{mathijssen2023culinary,norton2006computational}, and---in the context of the energy transition---latent thermal energy storage with phase-change materials (PCMs) \cite{vogel2019,li2019review,jouhara2020}. The latter is receiving recent industrial interest as it allows for the temporal mismatch between energy demand and supply to be mitigated, e.g. enabling the storage of the highly fluctuating solar or wind power in latent heat, and its later release when needed. 

The main objective of fundamental research on melting in turbulent flow must be a quantitative prediction of the melt rates as function of the size of the melting object, both for fixed and for freely-advected objects. This prediction is often sought for in the form of scaling laws, which are based on controlled experiments and observations on small scales and then extrapolated to much larger systems. However, such extrapolations only make sense if there are no transitions from one scaling relation to another one. Turbulent flows, however, can show transitions between different turbulent states. For example, in thermally driven turbulence, beyond a certain range of driving strengths the system undergoes a transition from classical turbulence, where the boundary layers are of laminar type, to so-called ultimate turbulence, where the boundary layers become turbulent, leading to enhanced heat transfer \cite{kra62,lohse2023,lohse2024}. For melting problems, such a transition would imply much larger melt rates above the transition. Indeed, the pioneering work by Machicoane {\it et al.}\ \cite{machicoane2013} on melting ice in turbulent flow identified a strong dependence of the melt rate on the degree of turbulence which corresponds to what we call ``ultimate melting". 

However, to actually observe the transitions from one regime to another one, i.e., from slower melt rates to faster ones, is intrinsically difficult, as many orders of magnitude must be covered to reliably identify the transition. In this study we achieve this anyhow and cover four orders of magnitude in the size of the melting object, namely by combining experiments of melting ice blocks of various sizes in statistically stationary, homogeneous, isotropic turbulent flows of different strengths with direct numerical simulations of melting objects in corresponding turbulent flows and with several reanalysed data sets from the literature \cite{machicoane2013,mccutchan2024}. In this sense our study can be seen as a meta-study: indeed, by this combination we do observe a clear transition between two quite different melting regimes, namely from slow or ``classical melting" on the small scales, to fast or ``ultimate melting'' on the large scales. This transition is due to a transition in the boundary layer around the melting object, namely from a laminar type boundary layer---allowing only for weak heat transport---to a turbulent boundary layer---with much enhanced heat transport. This transition occurs in the range of shear Reynolds numbers being of the order of one hundred, as expected from the analogous transition in wall-bounded shear flows and thermally driven turbulence (in so-called Rayleigh--B\'enard convection) \cite{lohse2023,lohse2024}. 
 
\begin{figure}
   \centering
   \includegraphics[width=0.6\textwidth]{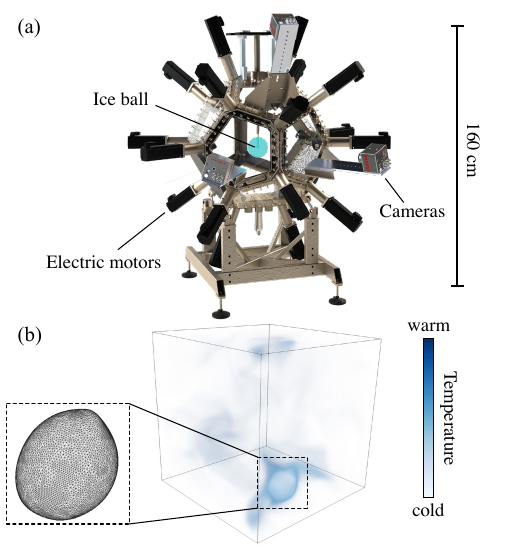}
    \caption{\textbf{Present experimental and numerical setups} (a) Experimental setup: twenty independently-controlled electric motors mounted on the vertices of a $\approx$\qty{210}{\liter} dodecahedral water tank drive propellers to generate HIT in the tank center. (b) Volume contours of instantaneous temperature in our present DNS at $\textit{Re}_{\lambda} \approx 50$, $\textit{Re}_{D_0} \approx 80$ for a Lagrangian ball at $50\%$ of its initial volume. The melting ball is represented by an immersed boundary \citep{zhong2025}, as visualized in the left panel.}
    \label{fig:setups}
\end{figure}

\begin{figure}
    \includegraphics[width=\textwidth]{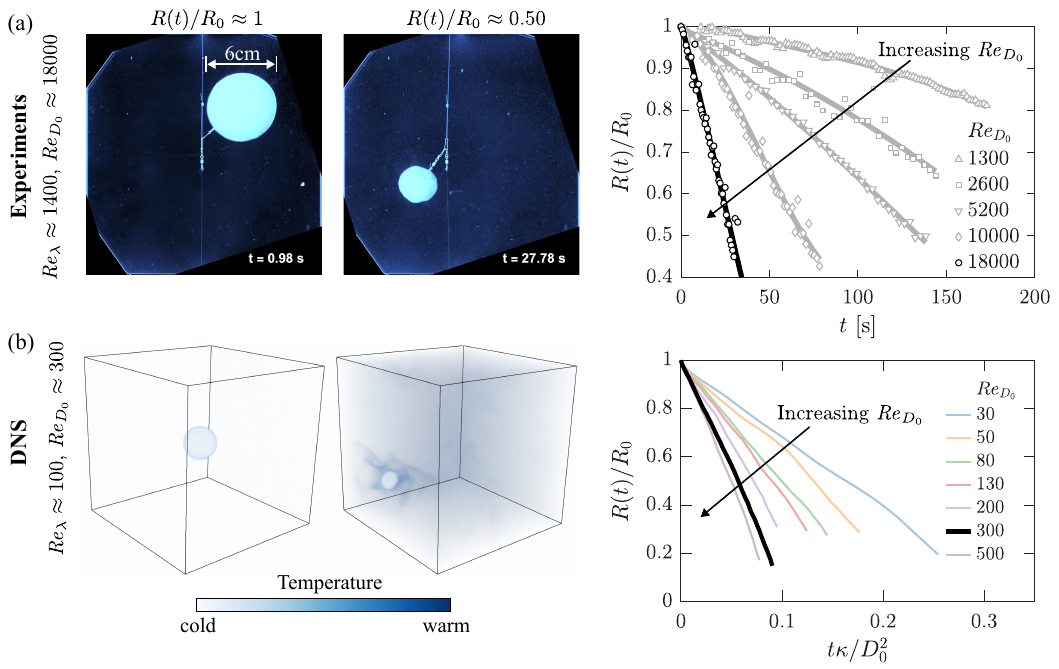}
    \caption{\textbf{Reynolds number controls melting rate of ice balls in both experiments and DNS}
    Snapshots of a freely-advected melting ice ball at two different times at fixed $\textit{Re}_\lambda$ and fixed $\textit{Re}_{D_0}$ (as noted) 
     and temporal evolution of the volume-equivalent radius for various $\textit{Re}_{D_0}$. The dark curve corresponds to the case shown 
     in the two left panels. 
    (a) Experimental results (b) DNS results.
    }
    \label{fig:time-traces}
\end{figure}

\section*{Results: Melting ice balls in experiments and simulations}
We perform experiments and simulations of ice balls melting in homogeneous isotropic turbulence, see figure \ref{fig:setups}. Sample movies visualizing the melting process in both our experiments and DNS are available as Supplementary Information. The details of both the experiments and DNSs can be found in \hyperref[sec:methods]{Methods}. We call balls that are fixed in space Eulerian, and balls that are advected by the turbulent flow Lagrangian. The main dimensionless control parameters of both experiments and DNSs are the Taylor--Reynolds number of the flow $\Rey_\lambda\equiv \sqrt{15 u_{\mathrm{rms}}^4/( \varepsilon \nu)}$, with $u_{\mathrm{rms}}$ the root-mean-square of the velocity's fluctuations, $\varepsilon$ the turbulent dissipation rate, $\nu$ the water's kinematic viscosity (which depends on temperature), and the initial ice ball Reynolds number $\Rey_{D_0} \equiv \mathcal{U}_{D_0} D_0/\nu$, where $\mathcal{U}_{D_0} \equiv (\varepsilon D_0)^{1/3}$ is the characteristic velocity of the turbulent flow at scale $D_0$ (the initial diameter of the ice ball). Further dimensionless control parameters of the system are the Prandtl number $\Pra \equiv \nu /\kappa$, where $\kappa$ is the thermal diffusivity of water (which depends on temperature), and the Stefan number $\Ste \equiv c_{p,w} \Delta T_{\mathrm{water}}/ \mathcal{L} $, where $\mathcal{L}$ is the latent heat of the ice, $c_{p,w}$ the specific heat of the water, and $\Delta T_{\mathrm{water}}$ the temperature of the surrounding water above the melting temperature.
 
Figure \ref{fig:time-traces}a shows two snapshots of a melting ice ball for the Lagrangian case for very strong turbulence ($Re_\lambda \approx 1400$) and an initial ice ball Reynolds number $\Rey_{D_0} \approx 18000$. Figure \ref{fig:time-traces}a, right, also displays the temporal evolution of the volume-equivalent radius $R(t)$ of the ice ball (relative to its initial size $R_0$) for that case $\Rey_{D_0} \approx 18000$ (dark curve) and various other cases with smaller $\Rey_{D_0}$ (light grey curves), for which the ice balls obviously melt more slowly. The analogous plots for DNS results are shown in figure \ref{fig:time-traces}b, though for a case of much weaker turbulence ($Re_\lambda \approx 100$; the largest $Re_\lambda$ available from DNS is about 200) and with much smaller $\Rey_{D_0} \approx 300$. 

Out of such temporal evolutions of the melting ice balls (either experimental or numerical) we can extract the mean melt rate $\langle \dot R \rangle$ (the average is over time) and its dependence on the control parameters. To allow for comparison between different experiments and simulations, the melt rate is expressed in dimensionless form as the Nusselt number $\Nu$, which is the dimensionless heat transfer from the water towards the ice (see \hyperref[sec:methods]{Methods} for a detailed derivation)
\begin{equation}\label{eq:connect}
   \Nu \approx - \frac{\rho_{\mathrm{ice}}}{\rho_{\mathrm{water}}}\frac\Pra\Ste \frac{\langle \dot{R} \rangle D_0}{\nu}\
\end{equation}

\begin{figure*}   
    \includegraphics[width=\textwidth]{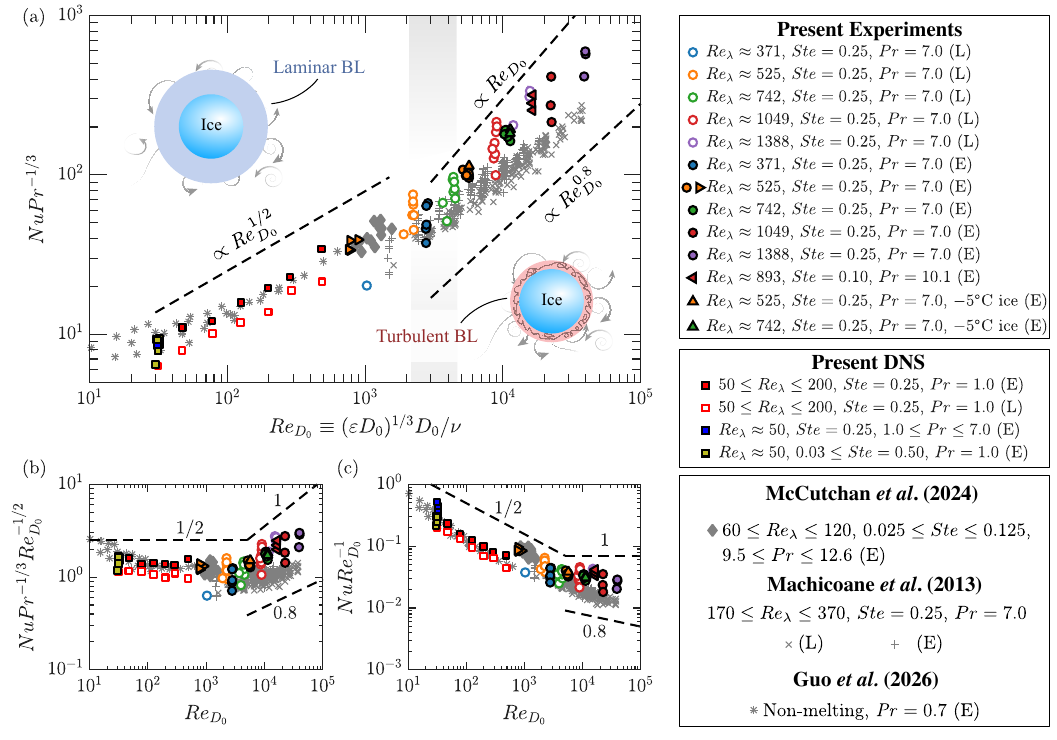}
    \caption{\textbf{A scaling transition from slow to fast melting}
     (a) Nusselt number compensated by $\textit{Pr}^{1/3}$ for all the presently considered experiments and DNS and data from the literature \citep{machicoane2013,mccutchan2024,guo2026}. The dashed lines indicate 
     $\textit{Nu} \propto \textit{Re}_{D_0}^{1/2}$, $\textit{Nu} \propto \textit{Re}_{D_0}^{0.8}$ and 
     $\textit{Nu} \propto \textit{Re}_{D_0}$ scaling relations. The symbol legend indicates the $\textit{Re}_\lambda$ considered, Stefan number $\textit{Ste}$, Prandtl number $\textit{Pr}$, and whether Eulerian (E) or Lagrangian (L) melting is considered. The grey-shaded region indicates a range where the laminar-to-turbulent boundary-layer transition takes place. In the two sketches, the boundary-layer thickness is not to scale. (b) Data in (a) compensated by $\textit{Re}_{D_0}^{1/2} \textit{Pr}^{1/3}$ to highlight the laminar boundary-layer scaling for $\textit{Re}_{D_0}\lesssim 4000$. (c) Data in (a) compensated by $\textit{Re}_{D_0}$ highlighting the ultimate regime scaling. The dashed lines in (b,c) indicate $\Nu \propto \textit{Re}_{D_0}^{1/2}$, $\Nu \propto \textit{Re}_{D_0}$, $\Nu \propto \textit{Re}_{D_0}^{0.8}$ scalings.
     }
     \label{fig:nu}
\end{figure*}

\noindent where $\rho_{\mathrm{ice}}/\rho_{\mathrm{water}} \approx 0.9$ is the ice-to-water density ratio. 

The results for $\Nu$ (compensated by $Pr^{1/3}$ for reasons to be explained later) vs $\Rey_{D_0}$ are shown in figure \ref{fig:nu}. The plot also contains (re-analysed) experimental data of Machicoane {\it et al.} \cite{machicoane2013}, McCutchan {\it et al.} \cite{mccutchan2024} (who varied $\Pra$ and $\Ste$), and Guo {\it et al.} \cite{guo2026} (for heat transfer without melting). 
Although the recent paper by Noto and Ulloa \cite{noto2026melting} also studies melting ice spheres, their interest is not on forced convection
but in natural convection (corresponding to $\Rey_{D_o}=0$); hence those data  cannot be included in the plot.
In the melting experiments, $\Rey_{D_0}$ varies by nearly two orders of magnitude, revealing a nearly linear dependence $\Nu \propto \Rey^\gamma_{D_0}$\cite{machicoane2013} with $0.8 \leq \gamma \leq 1.0$.
The numerical data in the plot reflect the considerable lower Reynolds numbers ($\Rey_{D_0} < 700 $), but note that the experimental and numerical data sets nearly overlap and show good agreement in the overall trend. 
From the experimental data, we cannot detect differences between the Eulerian and the Lagrangian cases; they are of similar magnitude as differences between repeated experiments for the same control parameters. 
The DNS data are more precise, and we can conclude that Eulerian melting is slightly faster than Lagrangian melting, due to the larger shear rates achieved along the melting object in the Eulerian case. 
But most importantly in the context of this work, four orders of magnitude in $\Rey_{D_0}$ are covered. 
This large range enables us to now clearly identify the existence of two different scaling regimes and the transition in between them, namely the above mentioned large $\Rey_{D_0}$ regime with $\Nu \propto \Rey^\gamma_{D_0}$ with ${0.8 \leq \gamma \leq 1.0}$ above the transition and a small $\Rey_{D_0}$ regime with $\Nu \propto \Rey^{1/2}_{D_0}$ below the transition. 

\section*{Discussion}
This latter scaling relation can be directly understood from the Prandtl--Blasius--Pohlhausen theory \cite{sch79,ahl09,lohse2024} of heat transfer through a laminar-type thermal boundary layer over a hot or cold plate, $\Nu \propto \Rey^{1/2} \Pra^{1/3}$ for $\Pra > 1$. This $\Pra$-dependence of $\Nu \propto \Pra^{1/3}$ is the reason why we compensate $\Nu$ with $\Pra^{1/3}$ in figure \ref{fig:nu}, in order to be able to compare results for different $\Pra$. For increasing Reynolds numbers, however, at some point the thermal boundary layer around the melting object can no longer be of laminar type, and undergoes a transition to a turbulent boundary layer of Prandtl--von K{\'a}rm{\'a}n type \cite{sch79,ahl09,lohse2024}. The nature of this transition is of non-normal, nonlinear type \cite{avi23}, i.e., there is no precise critical shear Reynolds number $\Rey_s$, but a range of $\Rey_s$ where the transition can happen, depending on the noise present in the system. For the canonical flat-plate boundary-layer one typically has $100 \leq \Rey_s \leq 500$  \cite{lohse2024}, whereas in our present case, the transition we observe occurs around $\Rey_{D_0} \approx 4000$, implying a shear Reynolds number of $\Rey_s \approx \Rey_{D_0}^{1/2} \approx 60$. The reasons why the observed onset of the transition is slightly earlier as compared to the classical flat-plate estimate are manifold: Besides inherent differences in geometry, well-understood effects which are known to promote boundary-layer receptivity (i.e. its sensitivity to noise) include \citep{sch79,schmid2001,drazin2004}: surface roughness \citep{dryden1953} (see for instance the classical textbook example of the dimpled golf ball to trigger early transition), freestream turbulence \citep{kendall1990}, unsteady flow effects \citep{obremski1967}, and 3D-effects or non-parallel flows \citep{klebanoff1962,saric1975}. All the effects mentioned are expected to be present in our melting configuration and account for the early $\Rey_s \approx 60$ transition onset.

Beyond the transition, i.e., for the case of a turbulent boundary layer around the melting object, the heat transfer scales as \cite{gro11,lohse2024} \be \Nu \propto \Rey \, \mathfrak{L}(Re) \label{karman} \ee for $\Pra >1$, as relevant for water. 
Here $\Rey$ is a system-size-scaled Reynolds number which here we identify with the Reynolds number $\Rey_{D_0}$ of the melting object and $\mathfrak{L}(Re)$ denotes logarithmic corrections with respect to $\Rey$. Over the range of relevant Reynolds numbers, relation (\ref{karman}) with its logarithmic correction is indistinguishable from an effective power law \cite{gro11} $\Nu \propto \Rey_{D_0}^{0.8} $, which is a common engineering correlation for turbulent heat transfer systems \citep{bej93,kays1993} and is consistent with the observed data in figure \ref{fig:nu}.

We note that the observed melting transition is analogous to the experimentally observed transition in heat transfer around a heated ball in air flows \cite{will2017}. Also in that case an increase in the Nusselt number beyond a certain Reynolds number range is achieved, analogous to the drag crisis for the drag force on the sphere.

There is yet another interesting aspect in the data of figure 3: while the dimensionless melt rate (as quantified by \textit{Nu}) clearly depends on $\textit{Re}_{D_0}$, the dependence of $\textit{Nu}$ with respect to the Taylor--Reynolds of the surrounding flow does not produce the same data collapse. For example, the green circles in figure 3 share the same $\textit{Re}_\lambda$, but differ significantly in $\textit{Re}_{D_0}$ and \textit{Nu}. This is because $\textit{Re}_\lambda$ does not distinguish between melting spheres of different diameters subject to the same ambient turbulent intensity.
Whereas $Re_\lambda$ characterises the full range of large- and small-eddies in the system, $\Rey_{D_0}$ characterises only the range from the melting object size to the smallest eddies. The fact that melting depends strongly on $\Rey_{D_0}$ but not on $\Rey_\lambda$, then, implies that only eddies of the size of the melting object and smaller are responsible for the melting process. The situation is similar as for droplets in turbulent flow, which according to the Kolmogorov--Hinze theory \cite{ni2024} become fragmented by eddies of the size of the droplet and smaller ones, but not by larger ones. 

\section*{Conclusions and Outlook}
In summary, by a meta-study of melting experiments and direct numerical simulations in turbulent flow over a very large range of four orders of magnitude in the Reynolds number of the melting object we have observed the transition from classical melting with $\Nu \propto \Rey_{D_0}^{1/2} $ to ultimate melting with $\Nu \propto \Rey_{D_0}^\gamma$ for $0.8 \leq \gamma \leq 1$, and interpret it as the transition from a laminar-type boundary layer to a turbulent boundary layer. The existence of this universal fluid dynamics transition in the melting process has profound implications for climate modelling: current parameterizations typically rely on simplified, fixed scaling relationships that do not capture the variability of processes across regimes and scales \cite{burgard_assessment_2022,yung_stratified_2025,jourdain_protocol_2020}, whereas we show that the dependence
of the melt rate
 on the Reynolds number cannot be obtained from simple extrapolations of scaling laws, but reflects transitions between different regimes.
Although the gap from laboratory experiments to geophysical scales is still far from being filled, identifying regime transitions and providing a sound understanding of the physics are key in safely extrapolating results of experiments from small scales to large ones---an ability which is greatly needed in the geophysical and industrial context.

\section*{Methods}\label{sec:methods}

\subsection*{Derivation of relation between melt rate and Nusselt number}
The starting point for the connection of $\langle \dot R \rangle $ and $\Nu$ is the integrated heat balance at the interface, given by the Stefan condition \citep{carslaw1959,alexiades1993,worster2000,davis2004}
\begin{equation}\label{eq:stefan1}
  -\mathcal{L} \rho_{\mathrm{ice}} \langle \dot{R} \rangle = \lambda_\mathrm{water} \bigg \langle \frac{\partial T}{\partial n} \bigg \rangle \bigg |_{\mathrm{water}} - \lambda_\mathrm{ice} \bigg \langle \frac{\partial T}{\partial n} \bigg \rangle \bigg |_{\mathrm{ice}}. 
\end{equation}
Here $\partial / \partial n \equiv \bnhat \cdot {\bf \nabla}$ denotes the gradient in the local interface-normal direction $\bnhat$, which points from the ice phase to the water phase and is evaluated at either the water-side of the interface ($|_{\mathrm{water}}$) or the ice-side ($|_{\mathrm{ice}}$), and $\lambda \equiv \rho c_p \kappa$ is the thermal conductivity of the respective phases. Here, the average $\langle \cdot \rangle$ refers to an average over the solid surface and over the melting duration. With the dimensionless heat transfer coefficients for the water-side convection, which is the Nusselt number 
\begin{equation}
    \Nu  \equiv \bigg \langle \frac{\partial T}{\partial n} \bigg \rangle \bigg |_{\mathrm{water}} \frac{D_0}{\Delta T_{\mathrm{water}}}, 
     \label{eq:nu-def}
\end{equation} 
and the ice-side conduction, which is the Biot number 
\begin{equation} 
    \textit{Bi} \equiv \bigg \langle \frac{\partial T}{\partial n} \bigg \rangle \bigg |_{\mathrm{ice}} \frac{D_0}{\Delta T_{\mathrm{ice}}},
       \label{eq:bi-def}
\end{equation}
we obtain the dimensionless form of the Stefan condition (\ref{eq:stefan1}), namely 
\begin{equation}\label{eq:balance_1}
    - \frac{\rho_{\mathrm{ice}}}{\rho_{\mathrm{water}}}\frac{\Pra}{\Ste} \frac{\langle \dot{R} \rangle D_0}{\nu} = \Nu - \frac{\lambda_{\mathrm{ice}}}{\lambda_{\mathrm{water}}} \, \frac{\Delta T_{\mathrm{ice}}}{\Delta T_{\mathrm{water}}}\, \textit{Bi}.
\end{equation}
This equation expresses a balance between the dimensionless melt rate $\langle \dot{R} \rangle D_0 / \nu$ and the sensible heat transfers on the water and ice side, given by the convective (water-side) heat transfer coefficient \Nu, and the solid-side heat transfer coefficient \textit{Bi}. 

The balance is typically simplified by recognizing that fluid convection can provide far more efficient heat transfer than conduction through a solid, such that $\Nu \gg \textit{Bi}$ \citep{dinniman2016}. Moreover, $\Delta T_{\mathrm{ice}}/\Delta T_{\mathrm{water}}$ and $\lambda_{\mathrm{ice}}/\lambda_{\mathrm{water}}$ are both $\mathcal{O}(1)$ prefactors in front of \textit{Bi} so the last term on the right-hand side of (\ref{eq:balance_1}) can be neglected. This then yields the desired relation between the measured dimensional melt rate $\langle \dot{R} \rangle$ of the ice block and the Nusselt number, namely eq.\ (\ref{eq:connect}).

\subsection*{Experimental setup}
Our experiments were carried out in a water tank in the shape of a regular dodecahedron, with the side measuring \qty{30}{\centi\meter}. The volume of the temperature-controlled water in the tank is approximately \qty{210}{\liter}. For most of the experiments, the temperature in the tank was kept at ambient temperature $T_\mathrm{amb}\approx$ \qty{20}{\degreeCelsius}. For the experiments at \textit{Ste} $=0.10$, \textit{Pr} $= 10.1$, the water was at \qty{7.5}{\celsius}$\pm$\qty{0.5}{\celsius}. At each of the 20 vertices of the dodecahedron 3D-printed toroidal propellers connected to electric motors drive the flow, each one with a maximum power of $\approx$\qty{1}{\kilo\watt}. With them we generate (approximately) homogeneous, isotropic, and statistically stationary turbulence, whose strength is quantified by its Taylor--Reynolds number $Re_\lambda$. We then put ice balls of three different diameters (\qty{3}{\centi\meter}, \qty{6}{\centi\meter}, and \qty{12}{\centi\meter}, corresponding to volumes of \qty{.01}{\liter}, \qty{.11}{\liter}, and \qty{0.90}{\liter}) into the flow. All the balls were made by freezing de-ionised (Milli-Q) water. The large balls were frozen with moulds of aluminium and PVC, while the small balls only with aluminium moulds. The Eulerian balls were frozen in an industrial freezer at \qty{-5}{\degreeCelsius} and then thermalised at \qty{-16}{\degreeCelsius}, except for the experiments marked as ``\qty{-5}{\degreeCelsius} ice'', where the balls were thermalised at \qty{-5}{\degreeCelsius}. The Lagrangian balls were shock-frozen using liquid nitrogen and the experiment was initiated quickly after unmoulding them. The balls were held either in an Eulerian way (i.e., at fixed position in the center of the flow) or in a (quasi)-Lagrangian way (i.e., by letting them freely advect in the bulk of the flow). We limited their motion with a vertically-stretched fishing wire connected to the ball and to the setup. This was done to have the ball advect only in the centre of the setup, where the flow is approximately homogeneously and isotropically
  turbulent. In addition, this procedure prevents collisions of the ice ball with the propellers and the walls.  For the Eulerian case, the size of the melting ice balls was measured by a single Nikon D850 camera with a Zeiss \qty{100}{\milli\meter} objective (resolution \qty{45}{\micro\meter\per\pixel}, \qty{2}{\fps}). For the Lagrangian case, we measured with three high-speed cameras (Photron AX200 with Nikon AF Nikkor 50 mm f/1.8D objective, resulting in a resolution of $\approx$ \qty{175}{\micro\meter\per\pixel}, operating at \qty{31.25}{\fps}), mounted on three adjacent windows of the dodecahedron. 

Experimentally, we measured the root-mean-square of the velocity fluctuations $u_\mathrm{rms}$ with precursor Lagrangian doppler velocimetry (LDV) measurements using a \qty{100}{\milli\meter} diameter PMMA sphere in a region of \qty{200}{\milli\meter} radius from the centre of the tank. The flow measurements were  not performed simultaneously  with the melting experiments, to avoid interference.
To obtain the dissipation rate $\varepsilon$, 
we applied the method described in van Buuren et al. \cite{buuren_twister_2025}.
  Our results show no preferential direction or angular variations in the statistics of the flow.

More details on the experiments, including details on the balls' 3D spherical fit method and detailed measurements of the turbulence in the tank, will be provided in a forthcoming publication.

\subsection*{Numerical setup}
Our direct numerical simulations (DNSs) were conducted by solving the underlying incompressible Navier--Stokes equations for the velocity field, together with the advection--diffusion equation for the temperature field. The computational domain was
 a cube with periodic boundary conditions applied in all three directions. Homogeneous isotropic turbulent flow conditions were
  sustained through a large-scale forcing which acted on the momentum field \citep{eswaran1988}. All simulations employed uniform grid spacing of size $\Delta$ for the Eulerian mesh, which was
   chosen to ensure proper resolution of the viscous Kolmogorov scale, $\eta$: $\Delta \lesssim 1.0 \eta$. Similarly, the grid spacing $\Delta$ was
    also chosen to ensure the solid geometry is adequately resolved, which was quantified by the number of grid points per initial sphere diameter $D_0/\Delta$. Our simulations employed $D_0 /\Delta \gtrsim 50$, lying well-above the typical $D_0 / \Delta \approx 30$ recommendation for interface-resolved simulations of particle-laden turbulence \citep{uhl05}.
The computational timestep, $\Delta t$, was
 kept well-below the Kolmogorov timescale, $\tau_\eta$: $\Delta t /\tau_\eta \lesssim 1/70$ ensuring stable time integration. To facilitate melting, 
 we used the method detailed in Zhong {\it et al.} \cite{zhong2025} which employs
  an immersed boundary framework to represent the melting ice and enforces the correct boundary conditions at the melting interface (i.e. the Stefan boundary condition \cite{worster2000}). The additional novelty of this framework is its ability to solve the Lagrangian melting case, where both the dynamics of the melting ice and fluid flow are two-way coupled and solved in tandem, allowing for the melting ice to be freely advected by the flow. Precise details on the implementation of the numerical method and its validation can be found in ref.\  \cite{zhong2025}. For each simulation, a precursor simulation was conducted to allow for a statistically steady flow field to develop,
   using the scheme as specified in \cite{eswaran1988}. The solid ice geometry remained fixed at the centre of the computational domain, i.e., no melting or solid motion occured during this period. The temperature field remained `frozen' during this precursor simulation, where the solid ice was
    initialised to be isothermal at the melting temperature $T_{\mathrm{melt}}$ and the ambient water was initialised with isothermal temperature $T_{\mathrm{amb}}$.  Further details on the full set of simulation runs presented and their employed and achieved parameter choices will be provided in a forthcoming publication.

\backmatter
\bmhead{Supplementary Information} This work has Supplementary Videos: two experiments ($\Rey_{D_0}\approx18000$, $\Rey_\lambda\approx1400$; and $\Rey_{D_0}\approx2600$, $\Rey_\lambda\approx500$) and two DNS ($\Rey_{D_0}\approx300$, $\Rey_\lambda\approx100$; and $\Rey_{D_0}\approx80$, $\Rey_\lambda\approx50$)

\bmhead{Acknlowledgements}
We thank Gert-Wim Bruggert, Martin Bos, and Thomas Zijlstra for technical support. We thank Romain Volk and Nathana\"el Machicoane for making the raw data of ref.\ \cite{machicoane2013} available to us. 
   
\section*{Declarations}
\begin{itemize}
    \item {\it Funding} This work was financially supported by the European Union (ERC, MeltDyn, 101040254 and ERC, MultiMelt, 101094492).
    \item {\it HPC} This work was carried out on the Dutch national e-infrastructure with the support of SURF Cooperative. We acknowledge the EuroHPC Joint Undertaking for awarding the project EHPC-REG-2023R03-178 access to the EuroHPC supercomputer Discoverer, hosted by Sofia Tech Park (Bulgaria). We also acknowledge the EuroHPC Joint Undertaking for awarding the project ID EHPC-EXT-2024E02-122 access to the MareNostrum V system hosted by the Barcelona Supercomputing Center (BSC), Spain.
    \item {\it Authors contributions} E.B., Y.Z., and S.G.H. performed the experiments; K.Z., C.J.H., and R.V. performed the DNS; D.L. conceived the idea. All authors edited the manuscript. 
    \item The authors declare no competing interests.
\end{itemize}


\begin{thebibliography}{63}
\ifx \bisbn   \undefined \def \bisbn  #1{ISBN #1}\fi
\ifx \binits  \undefined \def \binits#1{#1}\fi
\ifx \bauthor  \undefined \def \bauthor#1{#1}\fi
\ifx \batitle  \undefined \def \batitle#1{#1}\fi
\ifx \bjtitle  \undefined \def \bjtitle#1{#1}\fi
\ifx \bvolume  \undefined \def \bvolume#1{\textbf{#1}}\fi
\ifx \byear  \undefined \def \byear#1{#1}\fi
\ifx \bissue  \undefined \def \bissue#1{#1}\fi
\ifx \bfpage  \undefined \def \bfpage#1{#1}\fi
\ifx \blpage  \undefined \def \blpage #1{#1}\fi
\ifx \burl  \undefined \def \burl#1{\textsf{#1}}\fi
\ifx \doiurl  \undefined \def \doiurl#1{\url{https://doi.org/#1}}\fi
\ifx \betal  \undefined \def \betal{\textit{et al.}}\fi
\ifx \binstitute  \undefined \def \binstitute#1{#1}\fi
\ifx \binstitutionaled  \undefined \def \binstitutionaled#1{#1}\fi
\ifx \bctitle  \undefined \def \bctitle#1{#1}\fi
\ifx \beditor  \undefined \def \beditor#1{#1}\fi
\ifx \bpublisher  \undefined \def \bpublisher#1{#1}\fi
\ifx \bbtitle  \undefined \def \bbtitle#1{#1}\fi
\ifx \bedition  \undefined \def \bedition#1{#1}\fi
\ifx \bseriesno  \undefined \def \bseriesno#1{#1}\fi
\ifx \blocation  \undefined \def \blocation#1{#1}\fi
\ifx \bsertitle  \undefined \def \bsertitle#1{#1}\fi
\ifx \bsnm \undefined \def \bsnm#1{#1}\fi
\ifx \bsuffix \undefined \def \bsuffix#1{#1}\fi
\ifx \bparticle \undefined \def \bparticle#1{#1}\fi
\ifx \barticle \undefined \def \barticle#1{#1}\fi
\bibcommenthead
\ifx \bconfdate \undefined \def \bconfdate #1{#1}\fi
\ifx \botherref \undefined \def \botherref #1{#1}\fi
\ifx \url \undefined \def \url#1{\textsf{#1}}\fi
\ifx \bchapter \undefined \def \bchapter#1{#1}\fi
\ifx \bbook \undefined \def \bbook#1{#1}\fi
\ifx \bcomment \undefined \def \bcomment#1{#1}\fi
\ifx \oauthor \undefined \def \oauthor#1{#1}\fi
\ifx \citeauthoryear \undefined \def \citeauthoryear#1{#1}\fi
\ifx \endbibitem  \undefined \def \endbibitem {}\fi
\ifx \bconflocation  \undefined \def \bconflocation#1{#1}\fi
\ifx \arxivurl  \undefined \def \arxivurl#1{\textsf{#1}}\fi
\csname PreBibitemsHook\endcsname

\bibitem[\protect\citeauthoryear{Wang et~al.}{2021}]{met11081297}
\begin{botherref}
\oauthor{\bsnm{Wang}, \binits{Y.}},
\oauthor{\bsnm{Cao}, \binits{L.}},
\oauthor{\bsnm{Cheng}, \binits{Z.}},
\oauthor{\bsnm{Blanpain}, \binits{B.}},
\oauthor{\bsnm{Guo}, \binits{M.}}:
Mathematical methodology and metallurgical application of turbulence modelling: A review.
Metals
\textbf{11}(8)
(2021)
\end{botherref}
\endbibitem

\bibitem[\protect\citeauthoryear{Chakraborty}{2009}]{chakraborty2009}
\begin{barticle}
\bauthor{\bsnm{Chakraborty}, \binits{N.}}:
\batitle{The effects of turbulence on molten pool transport during melting and solidification processes in continuous conduction mode laser welding of copper--nickel dissimilar couple}.
\bjtitle{Appl. Therm. Eng.}
\bvolume{29}(\bissue{17-18}),
\bfpage{3618}--\blpage{3631}
(\byear{2009})
\end{barticle}
\endbibitem

\bibitem[\protect\citeauthoryear{Aboutalebi et~al.}{1995}]{aboutalebi1995coupled}
\begin{barticle}
\bauthor{\bsnm{Aboutalebi}, \binits{M.R.}},
\bauthor{\bsnm{Hasan}, \binits{M.}},
\bauthor{\bsnm{Guthrie}, \binits{R.}}:
\batitle{Coupled turbulent flow, heat, and solute transport in continuous casting processes}.
\bjtitle{Metall. Mater. Trans. b}
\bvolume{26}(\bissue{4}),
\bfpage{731}--\blpage{744}
(\byear{1995})
\end{barticle}
\endbibitem

\bibitem[\protect\citeauthoryear{Farrant}{1965}]{farrant1965}
\begin{barticle}
\bauthor{\bsnm{Farrant}, \binits{J.}}:
\batitle{Mechanism of cell damage during freezing and thawing and its prevention}.
\bjtitle{Nature}
\bvolume{205}(\bissue{4978}),
\bfpage{1284}--\blpage{1287}
(\byear{1965})
\end{barticle}
\endbibitem

\bibitem[\protect\citeauthoryear{Uhrig et~al.}{2022}]{uhrig2022}
\begin{barticle}
\bauthor{\bsnm{Uhrig}, \binits{M.}},
\bauthor{\bsnm{Ezquer}, \binits{F.}},
\bauthor{\bsnm{Ezquer}, \binits{M.}}:
\batitle{Improving cell recovery: freezing and thawing optimization of induced pluripotent stem cells}.
\bjtitle{Cells}
\bvolume{11}(\bissue{5}),
\bfpage{799}
(\byear{2022})
\end{barticle}
\endbibitem

\bibitem[\protect\citeauthoryear{Mathijssen et~al.}{2023}]{mathijssen2023culinary}
\begin{barticle}
\bauthor{\bsnm{Mathijssen}, \binits{A.J.}},
\bauthor{\bsnm{Lisicki}, \binits{M.}},
\bauthor{\bsnm{Prakash}, \binits{V.N.}},
\bauthor{\bsnm{Mossige}, \binits{E.J.}}:
\batitle{Culinary fluid mechanics and other currents in food science}.
\bjtitle{Rev. Mod. Phys.}
\bvolume{95}(\bissue{2}),
\bfpage{025004}
(\byear{2023})
\end{barticle}
\endbibitem

\bibitem[\protect\citeauthoryear{Norton and Sun}{2006}]{norton2006computational}
\begin{barticle}
\bauthor{\bsnm{Norton}, \binits{T.}},
\bauthor{\bsnm{Sun}, \binits{D.-W.}}:
\batitle{Computational fluid dynamics ({CFD})--an effective and efficient design and analysis tool for the food industry: {A} review}.
\bjtitle{Trends Food Sci. Technol.}
\bvolume{17}(\bissue{11}),
\bfpage{600}--\blpage{620}
(\byear{2006})
\end{barticle}
\endbibitem

\bibitem[\protect\citeauthoryear{Vogel and Thess}{2019}]{vogel2019}
\begin{barticle}
\bauthor{\bsnm{Vogel}, \binits{J.}},
\bauthor{\bsnm{Thess}, \binits{A.}}:
\batitle{Validation of a numerical model with a benchmark experiment for melting governed by natural convection in latent thermal energy storage}.
\bjtitle{Appl. Thermal Eng.}
\bvolume{148},
\bfpage{147}--\blpage{159}
(\byear{2019})
\end{barticle}
\endbibitem

\bibitem[\protect\citeauthoryear{Li et~al.}{2019}]{li2019review}
\begin{barticle}
\bauthor{\bsnm{Li}, \binits{Q.}},
\bauthor{\bsnm{Li}, \binits{C.}},
\bauthor{\bsnm{Du}, \binits{Z.}},
\bauthor{\bsnm{Jiang}, \binits{F.}},
\bauthor{\bsnm{Ding}, \binits{Y.}}:
\batitle{A review of performance investigation and enhancement of shell and tube thermal energy storage device containing molten salt based phase change materials for medium and high temperature applications}.
\bjtitle{Appl. Energy}
\bvolume{255},
\bfpage{113806}
(\byear{2019})
\end{barticle}
\endbibitem

\bibitem[\protect\citeauthoryear{Jouhara et~al.}{2020}]{jouhara2020}
\begin{barticle}
\bauthor{\bsnm{Jouhara}, \binits{H.}},
\bauthor{\bsnm{{\.Z}abnie{\'n}ska-G{\'o}ra}, \binits{A.}},
\bauthor{\bsnm{Khordehgah}, \binits{N.}},
\bauthor{\bsnm{Ahmad}, \binits{D.}},
\bauthor{\bsnm{Lipinski}, \binits{T.}}:
\batitle{Latent thermal energy storage technologies and applications: A review}.
\bjtitle{Int. J. Thermofluids}
\bvolume{5},
\bfpage{100039}
(\byear{2020})
\end{barticle}
\endbibitem

\bibitem[\protect\citeauthoryear{Hewitt}{2020}]{hewitt2020}
\begin{barticle}
\bauthor{\bsnm{Hewitt}, \binits{I.J.}}:
\batitle{Subglacial plumes}.
\bjtitle{Annu. Rev. Fluid Mech.}
\bvolume{52},
\bfpage{145}--\blpage{169}
(\byear{2020})
\end{barticle}
\endbibitem

\bibitem[\protect\citeauthoryear{Straneo and Cenedese}{2015}]{straneo2015}
\begin{barticle}
\bauthor{\bsnm{Straneo}, \binits{F.}},
\bauthor{\bsnm{Cenedese}, \binits{C.}}:
\batitle{The dynamics of greenland's glacial fjords and their role in climate}.
\bjtitle{Annu. Rev. Marine Sci.}
\bvolume{7},
\bfpage{89}--\blpage{112}
(\byear{2015})
\end{barticle}
\endbibitem

\bibitem[\protect\citeauthoryear{Cenedese and Straneo}{2023}]{cenedese2023}
\begin{barticle}
\bauthor{\bsnm{Cenedese}, \binits{C.}},
\bauthor{\bsnm{Straneo}, \binits{F.}}:
\batitle{Icebergs melting}.
\bjtitle{Annu. Rev. Fluid Mech.}
\bvolume{55}(\bissue{1}),
\bfpage{377}--\blpage{402}
(\byear{2023})
\end{barticle}
\endbibitem

\bibitem[\protect\citeauthoryear{Pritchard et~al.}{2012}]{pritchard2012}
\begin{barticle}
\bauthor{\bsnm{Pritchard}, \binits{H.}},
\bauthor{\bsnm{Ligtenberg}, \binits{S.R.}},
\bauthor{\bsnm{Fricker}, \binits{H.A.}},
\bauthor{\bsnm{Vaughan}, \binits{D.G.}},
\bauthor{\bsnm{Broeke}, \binits{M.R.}},
\bauthor{\bsnm{Padman}, \binits{L.}}:
\batitle{Antarctic ice-sheet loss driven by basal melting of ice shelves}.
\bjtitle{Nature}
\bvolume{484}(\bissue{7395}),
\bfpage{502}--\blpage{505}
(\byear{2012})
\end{barticle}
\endbibitem

\bibitem[\protect\citeauthoryear{Shepherd et~al.}{2012}]{shepherd2012}
\begin{barticle}
\bauthor{\bsnm{Shepherd}, \binits{A.}},
\bauthor{\bsnm{Ivins}, \binits{E.R.}},
\bauthor{\bsnm{A}, \binits{G.}},
\bauthor{\bsnm{Barletta}, \binits{V.R.}},
\bauthor{\bsnm{Bentley}, \binits{M.J.}},
\bauthor{\bsnm{Bettadpur}, \binits{S.}},
\bauthor{\bsnm{Briggs}, \binits{K.H.}},
\bauthor{\bsnm{Bromwich}, \binits{D.H.}},
\bauthor{\bsnm{Forsberg}, \binits{R.}},
\bauthor{\bsnm{Galin}, \binits{N.}}, \betal:
\batitle{A reconciled estimate of ice-sheet mass balance}.
\bjtitle{Science}
\bvolume{338}(\bissue{6111}),
\bfpage{1183}--\blpage{1189}
(\byear{2012})
\end{barticle}
\endbibitem

\bibitem[\protect\citeauthoryear{Stroeve et~al.}{2007}]{stroeve2007}
\begin{barticle}
\bauthor{\bsnm{Stroeve}, \binits{J.}},
\bauthor{\bsnm{Holland}, \binits{M.M.}},
\bauthor{\bsnm{Meier}, \binits{W.}},
\bauthor{\bsnm{Scambos}, \binits{T.}},
\bauthor{\bsnm{Serreze}, \binits{M.}}:
\batitle{Arctic sea ice decline: Faster than forecast}.
\bjtitle{Geophys. Res. Lett.}
\bvolume{34}(\bissue{9}),
\bfpage{09501}
(\byear{2007})
\end{barticle}
\endbibitem

\bibitem[\protect\citeauthoryear{Sutherland et~al.}{2019}]{sutherland2019}
\begin{barticle}
\bauthor{\bsnm{Sutherland}, \binits{D.}},
\bauthor{\bsnm{Jackson}, \binits{R.H.}},
\bauthor{\bsnm{Kienholz}, \binits{C.}},
\bauthor{\bsnm{Amundson}, \binits{J.M.}},
\bauthor{\bsnm{Dryer}, \binits{W.}},
\bauthor{\bsnm{Duncan}, \binits{D.}},
\bauthor{\bsnm{Eidam}, \binits{E.}},
\bauthor{\bsnm{Motyka}, \binits{R.}},
\bauthor{\bsnm{Nash}, \binits{J.}}:
\batitle{Direct observations of submarine melt and subsurface geometry at a tidewater glacier}.
\bjtitle{Science}
\bvolume{365}(\bissue{6451}),
\bfpage{369}--\blpage{374}
(\byear{2019})
\end{barticle}
\endbibitem

\bibitem[\protect\citeauthoryear{Buzzard}{2022}]{buzzard2022}
\begin{barticle}
\bauthor{\bsnm{Buzzard}, \binits{S.}}:
\batitle{The surface hydrology of {A}ntarctica's floating ice}.
\bjtitle{Phys. Today}
\bvolume{75}(\bissue{1}),
\bfpage{28}--\blpage{35}
(\byear{2022})
\end{barticle}
\endbibitem

\bibitem[\protect\citeauthoryear{Malyarenko et~al.}{2020}]{malyarenko2020}
\begin{barticle}
\bauthor{\bsnm{Malyarenko}, \binits{A.}},
\bauthor{\bsnm{Wells}, \binits{A.J.}},
\bauthor{\bsnm{Langhorne}, \binits{P.J.}},
\bauthor{\bsnm{Robinson}, \binits{N.J.}},
\bauthor{\bsnm{Williams}, \binits{M.J.}},
\bauthor{\bsnm{Nicholls}, \binits{K.W.}}:
\batitle{A synthesis of thermodynamic ablation at ice--ocean interfaces from theory, observations and models}.
\bjtitle{Ocean Modell.}
\bvolume{154},
\bfpage{101692}
(\byear{2020})
\end{barticle}
\endbibitem

\bibitem[\protect\citeauthoryear{Young et~al.}{2022}]{young2022}
\begin{barticle}
\bauthor{\bsnm{Young}, \binits{T.J.}},
\bauthor{\bsnm{Christoffersen}, \binits{P.}},
\bauthor{\bsnm{Bougamont}, \binits{M.}},
\bauthor{\bsnm{Tulaczyk}, \binits{S.M.}},
\bauthor{\bsnm{Hubbard}, \binits{B.}},
\bauthor{\bsnm{Mankoff}, \binits{K.D.}},
\bauthor{\bsnm{Nicholls}, \binits{K.W.}},
\bauthor{\bsnm{Stewart}, \binits{C.L.}}:
\batitle{Rapid basal melting of the {G}reenland {I}ce {S}heet from surface meltwater drainage}.
\bjtitle{Proc. Natl. Acad. Sci.}
\bvolume{119}(\bissue{10}),
\bfpage{2116036119}
(\byear{2022})
\end{barticle}
\endbibitem

\bibitem[\protect\citeauthoryear{Larter}{2022}]{larter2022}
\begin{barticle}
\bauthor{\bsnm{Larter}, \binits{R.D.}}:
\batitle{Basal melting, roughness and structural integrity of ice shelves}.
\bjtitle{Geophys. Res. Lett.}
\bvolume{49},
\bfpage{2021}--\blpage{097421}
(\byear{2022})
\end{barticle}
\endbibitem

\bibitem[\protect\citeauthoryear{Dauxois et~al.}{2021}]{dauxois2021}
\begin{barticle}
\bauthor{\bsnm{Dauxois}, \binits{T.}},
\bauthor{\bsnm{Peacock}, \binits{T.}},
\bauthor{\bsnm{Bauer}, \binits{P.}},
\bauthor{\bsnm{Caulfield}, \binits{C.P.}},
\bauthor{\bsnm{Cenedese}, \binits{C.}},
\bauthor{\bsnm{Gorl{\'e}}, \binits{C.}},
\bauthor{\bsnm{Haller}, \binits{G.}},
\bauthor{\bsnm{Ivey}, \binits{G.N.}},
\bauthor{\bsnm{Linden}, \binits{P.F.}},
\bauthor{\bsnm{Meiburg}, \binits{E.}},
\bauthor{\bsnm{Pinardi}, \binits{N.}},
\bauthor{\bsnm{Vriend}, \binits{N.M.}},
\bauthor{\bsnm{Woods}, \binits{A.W.}}:
\batitle{Confronting grand challenges in environmental fluid mechanics}.
\bjtitle{Phys. Rev. Fluids}
\bvolume{6}(\bissue{2}),
\bfpage{020501}
(\byear{2021})
\end{barticle}
\endbibitem

\bibitem[\protect\citeauthoryear{No{\"e}l et~al.}{2023}]{noel2023}
\begin{barticle}
\bauthor{\bsnm{No{\"e}l}, \binits{B.}},
\bauthor{\bsnm{Van~Wessem}, \binits{J.M.}},
\bauthor{\bsnm{Wouters}, \binits{B.}},
\bauthor{\bsnm{Trusel}, \binits{L.}},
\bauthor{\bsnm{Lhermitte}, \binits{S.}},
\bauthor{\bsnm{Van Den~Broeke}, \binits{M.R.}}:
\batitle{Higher antarctic ice sheet accumulation and surface melt rates revealed at 2 km resolution}.
\bjtitle{Nature Commun.}
\bvolume{14}(\bissue{1}),
\bfpage{7949}
(\byear{2023})
\end{barticle}
\endbibitem

\bibitem[\protect\citeauthoryear{Hanna et~al.}{2024}]{hanna2024}
\begin{barticle}
\bauthor{\bsnm{Hanna}, \binits{E.}},
\bauthor{\bsnm{Top{\'a}l}, \binits{D.}},
\bauthor{\bsnm{Box}, \binits{J.E.}},
\bauthor{\bsnm{Buzzard}, \binits{S.}},
\bauthor{\bsnm{Christie}, \binits{F.D.}},
\bauthor{\bsnm{Hvidberg}, \binits{C.}},
\bauthor{\bsnm{Morlighem}, \binits{M.}},
\bauthor{\bsnm{De~Santis}, \binits{L.}},
\bauthor{\bsnm{Silvano}, \binits{A.}},
\bauthor{\bsnm{Colleoni}, \binits{F.}}, \betal:
\batitle{Short-and long-term variability of the antarctic and greenland ice sheets}.
\bjtitle{Nature Reviews Earth \& Environment}
\bvolume{5}(\bissue{3}),
\bfpage{193}--\blpage{210}
(\byear{2024})
\end{barticle}
\endbibitem

\bibitem[\protect\citeauthoryear{Lucas et~al.}{2025}]{lucas2025}
\begin{botherref}
\oauthor{\bsnm{Lucas}, \binits{N.S.}},
\oauthor{\bsnm{Brearley}, \binits{J.A.}},
\oauthor{\bsnm{Hendry}, \binits{K.R.}},
\oauthor{\bsnm{Spira}, \binits{T.}},
\oauthor{\bsnm{Braakmann-Folgmann}, \binits{A.}},
\oauthor{\bsnm{Abrahamsen}, \binits{E.P.}},
\oauthor{\bsnm{Meredith}, \binits{M.P.}},
\oauthor{\bsnm{Tarling}, \binits{G.A.}}:
Giant iceberg meltwater increases upper-ocean stratification and vertical mixing.
Nature Geoscience,
1--8
(2025)
\end{botherref}
\endbibitem

\bibitem[\protect\citeauthoryear{Kraichnan}{1962}]{kra62}
\begin{barticle}
\bauthor{\bsnm{Kraichnan}, \binits{R.H.}}:
\batitle{Turbulent thermal convection at arbritrary {{Prandtl}} number}.
\bjtitle{Phys. Fluids}
\bvolume{5},
\bfpage{1374}--\blpage{1389}
(\byear{1962})
\end{barticle}
\endbibitem

\bibitem[\protect\citeauthoryear{Avila et~al.}{2023}]{avi23}
\begin{barticle}
\bauthor{\bsnm{Avila}, \binits{M.}},
\bauthor{\bsnm{Barkley}, \binits{D.}},
\bauthor{\bsnm{Hof}, \binits{B.}}:
\batitle{Transition to turbulence in pipe flow}.
\bjtitle{Annu. Rev. Fluid Mech.}
\bvolume{55},
\bfpage{575}--\blpage{602}
(\byear{2023})
\end{barticle}
\endbibitem

\bibitem[\protect\citeauthoryear{Lohse and Shishkina}{2023}]{lohse2023}
\begin{barticle}
\bauthor{\bsnm{Lohse}, \binits{D.}},
\bauthor{\bsnm{Shishkina}, \binits{O.}}:
\batitle{Ultimate turbulent thermal convection}.
\bjtitle{Phys. Today}
\bvolume{76}(\bissue{11}),
\bfpage{26}--\blpage{32}
(\byear{2023})
\end{barticle}
\endbibitem

\bibitem[\protect\citeauthoryear{Lohse and Shishkina}{2024}]{lohse2024}
\begin{barticle}
\bauthor{\bsnm{Lohse}, \binits{D.}},
\bauthor{\bsnm{Shishkina}, \binits{O.}}:
\batitle{{Ultimate Rayleigh--B\'enard turbulence}}.
\bjtitle{Rev. Mod. Phys.}
\bvolume{96},
\bfpage{035001}
(\byear{2024})
\end{barticle}
\endbibitem

\bibitem[\protect\citeauthoryear{Hester et~al.}{2021}]{hester2021}
\begin{barticle}
\bauthor{\bsnm{Hester}, \binits{E.W.}},
\bauthor{\bsnm{McConnochie}, \binits{C.D.}},
\bauthor{\bsnm{Cenedese}, \binits{C.}},
\bauthor{\bsnm{Couston}, \binits{L.-A.}},
\bauthor{\bsnm{Vasil}, \binits{G.}}:
\batitle{Aspect ratio affects iceberg melting}.
\bjtitle{Phys. Rev. Fluids}
\bvolume{6}(\bissue{2}),
\bfpage{023802}
(\byear{2021})
\end{barticle}
\endbibitem

\bibitem[\protect\citeauthoryear{Wells and Worster}{2008}]{wells_geophysical-scale_2008}
\begin{barticle}
\bauthor{\bsnm{Wells}, \binits{A.J.}},
\bauthor{\bsnm{Worster}, \binits{M.G.}}:
\batitle{A geophysical-scale model of vertical natural convection boundary layers}.
\bjtitle{J. Fluid Mech.}
\bvolume{609},
\bfpage{111}--\blpage{137}
(\byear{2008})
\end{barticle}
\endbibitem

\bibitem[\protect\citeauthoryear{Weady et~al.}{2022}]{weady2022}
\begin{barticle}
\bauthor{\bsnm{Weady}, \binits{S.}},
\bauthor{\bsnm{Tong}, \binits{J.}},
\bauthor{\bsnm{Zidovska}, \binits{A.}},
\bauthor{\bsnm{Ristroph}, \binits{L.}}:
\batitle{Anomalous convective flows carve pinnacles and scallops in melting ice}.
\bjtitle{Phys. Rev. Lett.}
\bvolume{128},
\bfpage{044502}
(\byear{2022})
\end{barticle}
\endbibitem

\bibitem[\protect\citeauthoryear{Rubinstein}{1971}]{rubinstein1971}
\begin{bbook}
\bauthor{\bsnm{Rubinstein}, \binits{L.}}:
\bbtitle{The {{Stefan}} Problem}.
\bpublisher{American Mathematical Society},
\blocation{Providence}
(\byear{1971})
\end{bbook}
\endbibitem

\bibitem[\protect\citeauthoryear{Machicoane et~al.}{2013}]{machicoane2013}
\begin{barticle}
\bauthor{\bsnm{Machicoane}, \binits{N.}},
\bauthor{\bsnm{Bonaventure}, \binits{J.}},
\bauthor{\bsnm{Volk}, \binits{R.}}:
\batitle{Melting dynamics of large ice balls in a turbulent swirling flow}.
\bjtitle{Phys. Fluids}
\bvolume{25}(\bissue{12}),
\bfpage{125101}
(\byear{2013})
\end{barticle}
\endbibitem

\bibitem[\protect\citeauthoryear{McCutchan et~al.}{2024}]{mccutchan2024}
\begin{barticle}
\bauthor{\bsnm{McCutchan}, \binits{A.L.}},
\bauthor{\bsnm{Meyer}, \binits{C.R.}},
\bauthor{\bsnm{Johnson}, \binits{B.A.}}:
\batitle{Enhancement of ice melting in isotropic turbulence}.
\bjtitle{Phys. Rev. Fluids}
\bvolume{9}(\bissue{7}),
\bfpage{074601}
(\byear{2024})
\end{barticle}
\endbibitem

\bibitem[\protect\citeauthoryear{Zhong et~al.}{2025}]{zhong2025}
\begin{barticle}
\bauthor{\bsnm{Zhong}, \binits{K.}},
\bauthor{\bsnm{Howland}, \binits{C.J.}},
\bauthor{\bsnm{Lohse}, \binits{D.}},
\bauthor{\bsnm{Verzicco}, \binits{R.}}:
\batitle{A front-tracking immersed-boundary framework for simulating {L}agrangian melting problems}.
\bjtitle{J.~Comput. Phys.}
\bvolume{525},
\bfpage{113762}
(\byear{2025})
\end{barticle}
\endbibitem

\bibitem[\protect\citeauthoryear{Guo et~al.}{2026}]{guo2026}
\begin{barticle}
\bauthor{\bsnm{Guo}, \binits{H.}},
\bauthor{\bsnm{Zhang}, \binits{X.}},
\bauthor{\bsnm{Feng}, \binits{L.}},
\bauthor{\bsnm{Wu}, \binits{Y.}},
\bauthor{\bsnm{Wu}, \binits{Y.}}:
\batitle{Experimental {N}usselt number correlations for heat transfer of a single spherical particle in turbulent flow}.
\bjtitle{Int. J. Heat Mass Transf.}
\bvolume{256},
\bfpage{127939}
(\byear{2026})
\end{barticle}
\endbibitem

\bibitem[\protect\citeauthoryear{Noto and Ulloa}{2026}]{noto2026melting}
\begin{barticle}
\bauthor{\bsnm{Noto}, \binits{D.}},
\bauthor{\bsnm{Ulloa}, \binits{H.N.}}:
\batitle{Melting dynamics of freely floating ice in calm waters}.
\bjtitle{Science Advances}
\bvolume{12}(\bissue{5}),
\bfpage{3529}
(\byear{2026})
\end{barticle}
\endbibitem

\bibitem[\protect\citeauthoryear{Schlichting}{1979}]{sch79}
\begin{bbook}
\bauthor{\bsnm{Schlichting}, \binits{H.}}:
\bbtitle{Boundary Layer Theory},
\bedition{7th} edn.
\bpublisher{McGraw Hill},
\blocation{New York}
(\byear{1979})
\end{bbook}
\endbibitem

\bibitem[\protect\citeauthoryear{Ahlers et~al.}{2009}]{ahl09}
\begin{barticle}
\bauthor{\bsnm{Ahlers}, \binits{G.}},
\bauthor{\bsnm{Grossmann}, \binits{S.}},
\bauthor{\bsnm{Lohse}, \binits{D.}}:
\batitle{Heat transfer and large scale dynamics in turbulent {{{{Rayleigh--B\'enard}}}} convection}.
\bjtitle{Rev. Mod. Phys.}
\bvolume{81},
\bfpage{503}
(\byear{2009})
\end{barticle}
\endbibitem

\bibitem[\protect\citeauthoryear{Schmid and Henningson}{2001}]{schmid2001}
\begin{bbook}
\bauthor{\bsnm{Schmid}, \binits{P.J.}},
\bauthor{\bsnm{Henningson}, \binits{D.S.}}:
\bbtitle{Stability and Transition in Shear Flows},
\bedition{1st} edn.
\bpublisher{Springer},
\blocation{New York}
(\byear{2001})
\end{bbook}
\endbibitem

\bibitem[\protect\citeauthoryear{Drazin and Reid}{2004}]{drazin2004}
\begin{bbook}
\bauthor{\bsnm{Drazin}, \binits{P.G.}},
\bauthor{\bsnm{Reid}, \binits{W.H.}}:
\bbtitle{Hydrodynamic Stability},
\bedition{2nd} edn.
\bpublisher{Cambridge University Press},
\blocation{Cambridge}
(\byear{2004})
\end{bbook}
\endbibitem

\bibitem[\protect\citeauthoryear{Dryden}{1953}]{dryden1953}
\begin{barticle}
\bauthor{\bsnm{Dryden}, \binits{H.L.}}:
\batitle{Review of published data on the effect of roughness on transition from laminar to turbulent flow}.
\bjtitle{J. Aero. Sci.}
\bvolume{20}(\bissue{7}),
\bfpage{477}--\blpage{482}
(\byear{1953})
\end{barticle}
\endbibitem

\bibitem[\protect\citeauthoryear{Kendall}{1990}]{kendall1990}
\begin{botherref}
\oauthor{\bsnm{Kendall}, \binits{J.}}:
Boundary layer receptivity to freestream turbulence.
AIAA Paper,
90--1504
(1990)
\end{botherref}
\endbibitem

\bibitem[\protect\citeauthoryear{Obremski and Fejer}{1967}]{obremski1967}
\begin{barticle}
\bauthor{\bsnm{Obremski}, \binits{H.J.}},
\bauthor{\bsnm{Fejer}, \binits{A.A.}}:
\batitle{Transition in oscillating boundary layer flows}.
\bjtitle{J. Fluid Mech.}
\bvolume{29}(\bissue{1}),
\bfpage{93}--\blpage{111}
(\byear{1967})
\end{barticle}
\endbibitem

\bibitem[\protect\citeauthoryear{Klebanoff et~al.}{1962}]{klebanoff1962}
\begin{barticle}
\bauthor{\bsnm{Klebanoff}, \binits{P.S.}},
\bauthor{\bsnm{Tidstrom}, \binits{K.D.}},
\bauthor{\bsnm{Sargent}, \binits{L.M.}}:
\batitle{The three-dimensional nature of boundary-layer instability}.
\bjtitle{J. Fluid Mech.}
\bvolume{12}(\bissue{1}),
\bfpage{1}--\blpage{34}
(\byear{1962})
\end{barticle}
\endbibitem

\bibitem[\protect\citeauthoryear{Saric and Nayfeh}{1975}]{saric1975}
\begin{barticle}
\bauthor{\bsnm{Saric}, \binits{W.S.}},
\bauthor{\bsnm{Nayfeh}, \binits{A.H.}}:
\batitle{Nonparallel stability of boundary-layer flows}.
\bjtitle{Phys. Fluids}
\bvolume{18}(\bissue{8}),
\bfpage{945}--\blpage{950}
(\byear{1975})
\end{barticle}
\endbibitem

\bibitem[\protect\citeauthoryear{Grossmann and Lohse}{2011}]{gro11}
\begin{barticle}
\bauthor{\bsnm{Grossmann}, \binits{S.}},
\bauthor{\bsnm{Lohse}, \binits{D.}}:
\batitle{Multiple scaling in the ultimate regime of thermal convection}.
\bjtitle{Phys. Fluids}
\bvolume{23},
\bfpage{045108}
(\byear{2011})
\end{barticle}
\endbibitem

\bibitem[\protect\citeauthoryear{Bejan}{1993}]{bej93}
\begin{bbook}
\bauthor{\bsnm{Bejan}, \binits{A.}}:
\bbtitle{Heat Transfer}.
\bpublisher{John Wiley and Sons. Inc.},
\blocation{New York}
(\byear{1993})
\end{bbook}
\endbibitem

\bibitem[\protect\citeauthoryear{Kays and Crawford}{1993}]{kays1993}
\begin{bbook}
\bauthor{\bsnm{Kays}, \binits{W.M.}},
\bauthor{\bsnm{Crawford}, \binits{M.E.}}:
\bbtitle{Convective Heat and Mass Transfer},
\bedition{3rd} edn.
\bpublisher{McGraw-Hill},
\blocation{New York}
(\byear{1993})
\end{bbook}
\endbibitem

\bibitem[\protect\citeauthoryear{Will et~al.}{2017}]{will2017}
\begin{barticle}
\bauthor{\bsnm{Will}, \binits{J.B.}},
\bauthor{\bsnm{Kruyt}, \binits{N.P.}},
\bauthor{\bsnm{Venner}, \binits{C.H.}}:
\batitle{An experimental study of forced convective heat transfer from smooth, solid spheres}.
\bjtitle{Int. J. Heat Mass Transf.}
\bvolume{109},
\bfpage{1059}--\blpage{1067}
(\byear{2017})
\end{barticle}
\endbibitem

\bibitem[\protect\citeauthoryear{Ni}{2024}]{ni2024}
\begin{barticle}
\bauthor{\bsnm{Ni}, \binits{R.}}:
\batitle{Deformation and breakup of bubbles and drops in turbulence}.
\bjtitle{Annu. Rev. Fluid Mech.}
\bvolume{56}(\bissue{1}),
\bfpage{319}--\blpage{347}
(\byear{2024})
\end{barticle}
\endbibitem

\bibitem[\protect\citeauthoryear{Burgard et~al.}{2022}]{burgard_assessment_2022}
\begin{barticle}
\bauthor{\bsnm{Burgard}, \binits{C.}},
\bauthor{\bsnm{Jourdain}, \binits{N.C.}},
\bauthor{\bsnm{Reese}, \binits{R.}},
\bauthor{\bsnm{Jenkins}, \binits{A.}},
\bauthor{\bsnm{Mathiot}, \binits{P.}}:
\batitle{An assessment of basal melt parameterisations for {Antarctic} ice shelves}.
\bjtitle{The Cryosphere}
\bvolume{16}(\bissue{12}),
\bfpage{4931}--\blpage{4975}
(\byear{2022})
\doiurl{10.5194/tc-16-4931-2022} .
Accessed 2026-04-30
\end{barticle}
\endbibitem

\bibitem[\protect\citeauthoryear{Yung et~al.}{2025}]{yung_stratified_2025}
\begin{barticle}
\bauthor{\bsnm{Yung}, \binits{C.K.}},
\bauthor{\bsnm{Rosevear}, \binits{M.G.}},
\bauthor{\bsnm{Morrison}, \binits{A.K.}},
\bauthor{\bsnm{Hogg}, \binits{A.M.}},
\bauthor{\bsnm{Nakayama}, \binits{Y.}}:
\batitle{Stratified suppression of turbulence in an ice shelf\\ basal melt parameterisation}.
\bjtitle{The Cryosphere}
\bvolume{19}(\bissue{11}),
\bfpage{5827}--\blpage{5861}
(\byear{2025})
\doiurl{10.5194/tc-19-5827-2025}
\end{barticle}
\endbibitem

\bibitem[\protect\citeauthoryear{Jourdain et~al.}{2020}]{jourdain_protocol_2020}
\begin{barticle}
\bauthor{\bsnm{Jourdain}, \binits{N.C.}},
\bauthor{\bsnm{Asay-Davis}, \binits{X.}},
\bauthor{\bsnm{Hattermann}, \binits{T.}},
\bauthor{\bsnm{Straneo}, \binits{F.}},
\bauthor{\bsnm{Seroussi}, \binits{H.}},
\bauthor{\bsnm{Little}, \binits{C.M.}},
\bauthor{\bsnm{Nowicki}, \binits{S.}}:
\batitle{A protocol for calculating basal melt rates in the {ISMIP6} {Antarctic} ice sheet projections}.
\bjtitle{The Cryosphere}
\bvolume{14}(\bissue{9}),
\bfpage{3111}--\blpage{3134}
(\byear{2020})
\doiurl{10.5194/tc-14-3111-2020} .
Accessed 2026-04-30
\end{barticle}
\endbibitem

\bibitem[\protect\citeauthoryear{Carslaw and Jaeger}{1959}]{carslaw1959}
\begin{bbook}
\bauthor{\bsnm{Carslaw}, \binits{H.S.}},
\bauthor{\bsnm{Jaeger}, \binits{J.C.}}:
\bbtitle{Conduction of {Heat} in {Solids}}.
\bpublisher{Oxford University Press},
\blocation{Oxford}
(\byear{1959})
\end{bbook}
\endbibitem

\bibitem[\protect\citeauthoryear{Alexiades and Solomon}{1993}]{alexiades1993}
\begin{bbook}
\bauthor{\bsnm{Alexiades}, \binits{V.}},
\bauthor{\bsnm{Solomon}, \binits{A.D.}}:
\bbtitle{Mathematical Modeling of Melting and Freezing Processes}.
\bpublisher{Taylor \& Francis},
\blocation{London}
(\byear{1993}).
\doiurl{10.1201/9780203749449}
\end{bbook}
\endbibitem

\bibitem[\protect\citeauthoryear{Worster}{2000}]{worster2000}
\begin{bchapter}
\bauthor{\bsnm{Worster}, \binits{M.G.}}:
\bctitle{Solidification of fluids}.
In: \bbtitle{Perspectives in Fluid Dynamics},
pp. \bfpage{393}--\blpage{446}.
\bpublisher{Cambridge University Press},
\blocation{Cambridge}
(\byear{2000})
\end{bchapter}
\endbibitem

\bibitem[\protect\citeauthoryear{Davis}{2004}]{davis2004}
\begin{bbook}
\bauthor{\bsnm{Davis}, \binits{S.H.}}:
\bbtitle{Theory of Solidification}.
\bpublisher{Cambridge University Press},
\blocation{Cambridge}
(\byear{2004})
\end{bbook}
\endbibitem

\bibitem[\protect\citeauthoryear{Dinniman et~al.}{2016}]{dinniman2016}
\begin{barticle}
\bauthor{\bsnm{Dinniman}, \binits{M.S.}},
\bauthor{\bsnm{Asay-Davis}, \binits{X.S.}},
\bauthor{\bsnm{Galton-Fenzi}, \binits{B.K.}},
\bauthor{\bsnm{Holland}, \binits{P.R.}},
\bauthor{\bsnm{Jenkins}, \binits{A.}},
\bauthor{\bsnm{Timmermann}, \binits{R.}}:
\batitle{Modeling ice shelf/ocean interaction in antarctica: A review}.
\bjtitle{Oceanography}
\bvolume{29}(\bissue{4}),
\bfpage{144}--\blpage{153}
(\byear{2016})
\doiurl{10.5670/oceanog.2016.106}
\end{barticle}
\endbibitem

\bibitem[\protect\citeauthoryear{{van Buuren} et~al.}{2025}]{buuren_twister_2025}
\begin{barticle}
\bauthor{\bsnm{{van Buuren}}, \binits{D.}},
\bauthor{\bsnm{{van Gils}}, \binits{D.P.M.}},
\bauthor{\bsnm{Bruggert}, \binits{G.-W.}},
\bauthor{\bsnm{Krug}, \binits{D.}}:
\batitle{{TWISTER} ({Twente} water injection system for turbulence experimental research): a jet array in the {Twente} water tunnel for generating strong turbulence using four-dimensional gradient noise}.
\bjtitle{Exp Fluids}
\bvolume{66}(\bissue{10}),
\bfpage{184}
(\byear{2025})
\end{barticle}
\endbibitem

\bibitem[\protect\citeauthoryear{Eswaran and Pope}{1988}]{eswaran1988}
\begin{barticle}
\bauthor{\bsnm{Eswaran}, \binits{V.}},
\bauthor{\bsnm{Pope}, \binits{S.B.}}:
\batitle{An examination of forcing in direct numerical simulations of turbulence}.
\bjtitle{Comput. Fluids}
\bvolume{16}(\bissue{3}),
\bfpage{257}--\blpage{278}
(\byear{1988})
\end{barticle}
\endbibitem

\bibitem[\protect\citeauthoryear{Uhlmann}{2005}]{uhl05}
\begin{barticle}
\bauthor{\bsnm{Uhlmann}, \binits{M.}}:
\batitle{An immersed boundary method with direct forcing for the simulation of particulate flows}.
\bjtitle{J. Comp. Phys}
\bvolume{209},
\bfpage{448}--\blpage{476}
(\byear{2005})
\end{barticle}
\endbibitem

\end{thebibliography}
\end{document}